\documentclass[conference]{IEEEtran}
\usepackage{cite}
\ifCLASSINFOpdf
\usepackage[pdftex]{graphicx}
\else
\fi
\usepackage[cmex10]{amsmath}
\usepackage{amsfonts}
\usepackage{relsize}

\newtheorem{Proposition}{Proposition}

\newtheorem{Algorithm}{Algorithm}

\usepackage{wrapfig}
\usepackage{algorithmic}
\usepackage{array}
\usepackage{mdwmath}
\usepackage{mdwtab}
\usepackage{color}
\definecolor{orange}{rgb}{1,0.7,0}

\newcommand{\E}{\ensuremath{\operatorname{\mathrm{E}}}}
\newcommand{\var}{\ensuremath{\operatorname{\mathrm{Var}}}}

\usepackage{subcaption,graphicx,algorithm,algorithmic,url,setspace}
\let\Algorithm\algorithm
\renewcommand\algorithm[1][]{\Algorithm[#1]\setstretch{1.3}}
\begin{document}
\title{A Power Variance Test for Nonstationarity in Complex-Valued Signals}
\author{\IEEEauthorblockN{Thomas E. Bartlett\\ and Adam M. Sykulski\\ and Sofia C. Olhede}
\IEEEauthorblockA{Department of Statistical Science\\
University College London, UK\\
Email: thomas.bartlett@ucl.ac.uk}
\and
\IEEEauthorblockN{Jonathan M. Lilly\\ and Jeffrey J. Early}
\IEEEauthorblockA{NorthWest Research Associates\\
Seattle, USA}}
\maketitle

\begin{abstract}
We propose a novel algorithm for testing the hypothesis of nonstationarity in complex-valued signals. The implementation uses both the bootstrap and the Fast Fourier Transform such that the algorithm can be efficiently implemented in $\mathcal{O}(N\text{log}N)$ time, where $N$ is the length of the observed signal. The test procedure examines the second-order structure and contrasts the observed power variance---€"i.e. the variability of the instantaneous variance over time---€"with the expected characteristics of stationary signals generated via the bootstrap method. Our algorithmic procedure is capable of learning different types of nonstationarity, such as jumps or strong sinusoidal components. We illustrate the utility of our test and algorithm through application to turbulent flow data from fluid dynamics.
\end{abstract}

\IEEEpeerreviewmaketitle

\copyright\ 2015 IEEE. Personal use of this material is permitted. Permission from IEEE must be obtained for all other uses, in any current or future media, including reprinting/republishing this material for advertising or promotional purposes, creating new collective works, for resale or redistribution to servers or lists, or reuse of any copyrighted component of this work in other works.

\vspace{-1mm}
\section{Introduction}
In this paper we introduce a new automated method of detecting nonstationarity in observed time signals. Identifying and dealing with nonstationarity is a problem of fundamental importance in machine learning~\cite{sugiyama2012machine,robinson2010learning,michaud1998learning}. Nonstationarity can be hard to detect, because while there is only one way for a signal to be stationary, there are many ways for this assumption to be violated~\cite{priestley1988non}. Our approach is based on examining the variance of the observed signal. As we assume the signal is zero mean, or that the mean is removed, this is the simplest way that the signal may depart from stationarity.

We focus our attention on bivariate signals, that are an important class of observations. Such data are commonly represented as complex-valued signals, as discussed in~\cite{schreier2010statistical}, and as performed in numerous applications including fMRI~\cite{rowe2005modeling}, blood flow~ \cite{brands1997radio}, neural networks~ \cite{mandic2009complex}, oceanography~\cite{sykulski2015lagrangian}, seismology~\cite{samson1977matrix}, and meteorology~\cite{gonella1972rotary}. The complex-valued representation is particularly useful for modelling trajectories of fluid particles in oceaongraphy and related applications~\cite{lilly2011extracting}, also referred to by the name ``Lagrangian data." In Fig.~1 we display three trajectories of particles propagated in a simulation of forced-dissipative two-dimensional fluid turbulence. These trajectories exhibit various degrees and types of nonstationarity, and we will test our procedures on this data in this paper.

\begin{figure*}[ht!]
\centering
\vspace{-4mm}
\begin{subfigure}{0.32\textwidth}
\includegraphics[width=0.98\textwidth]{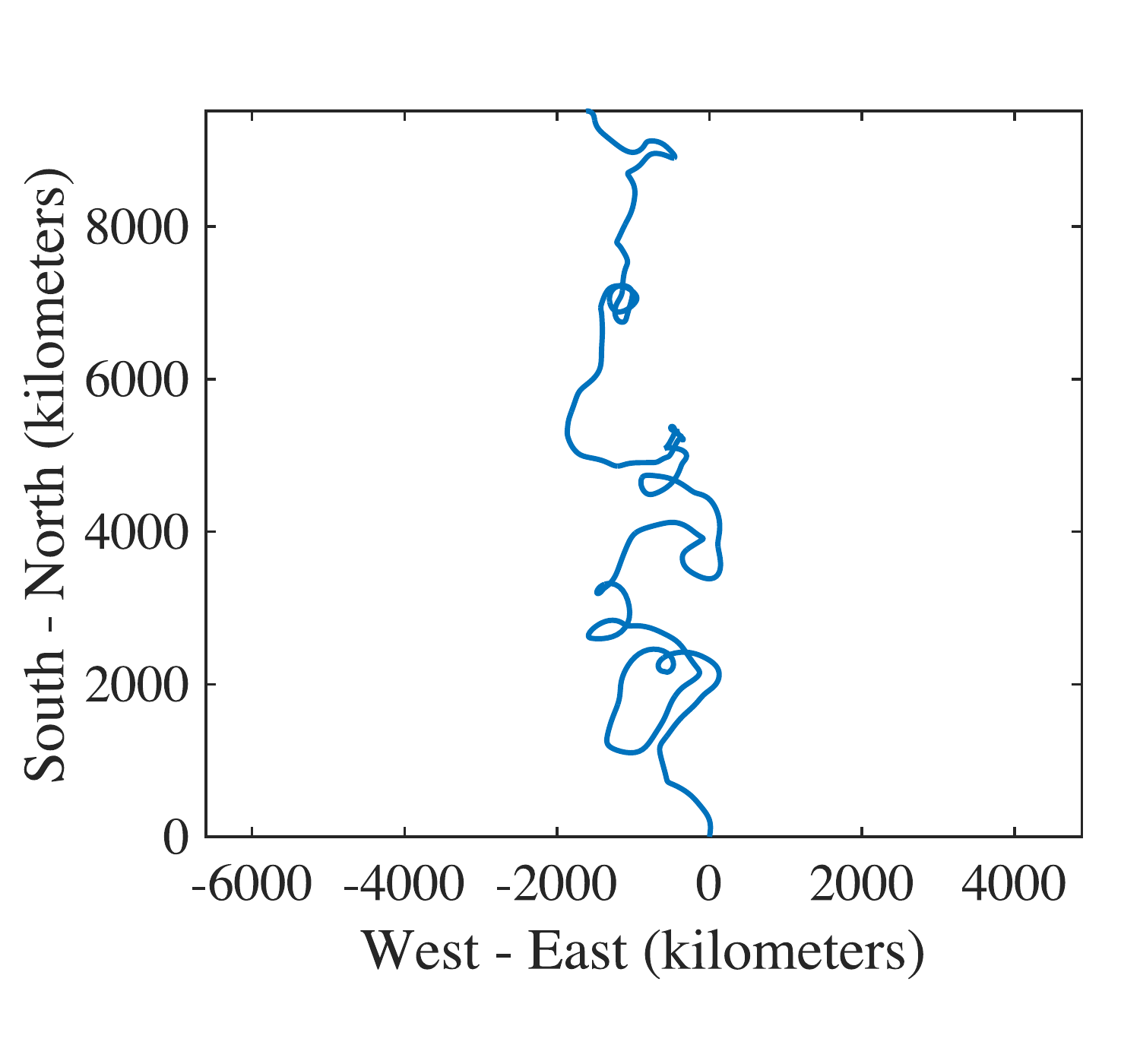}
\vspace{-4mm}
\caption{}\label{fig:1a}
\end{subfigure}
\hspace*{-2mm} 
\vspace{-4mm}
\begin{subfigure}{0.32\textwidth}
\includegraphics[width=0.98\textwidth]{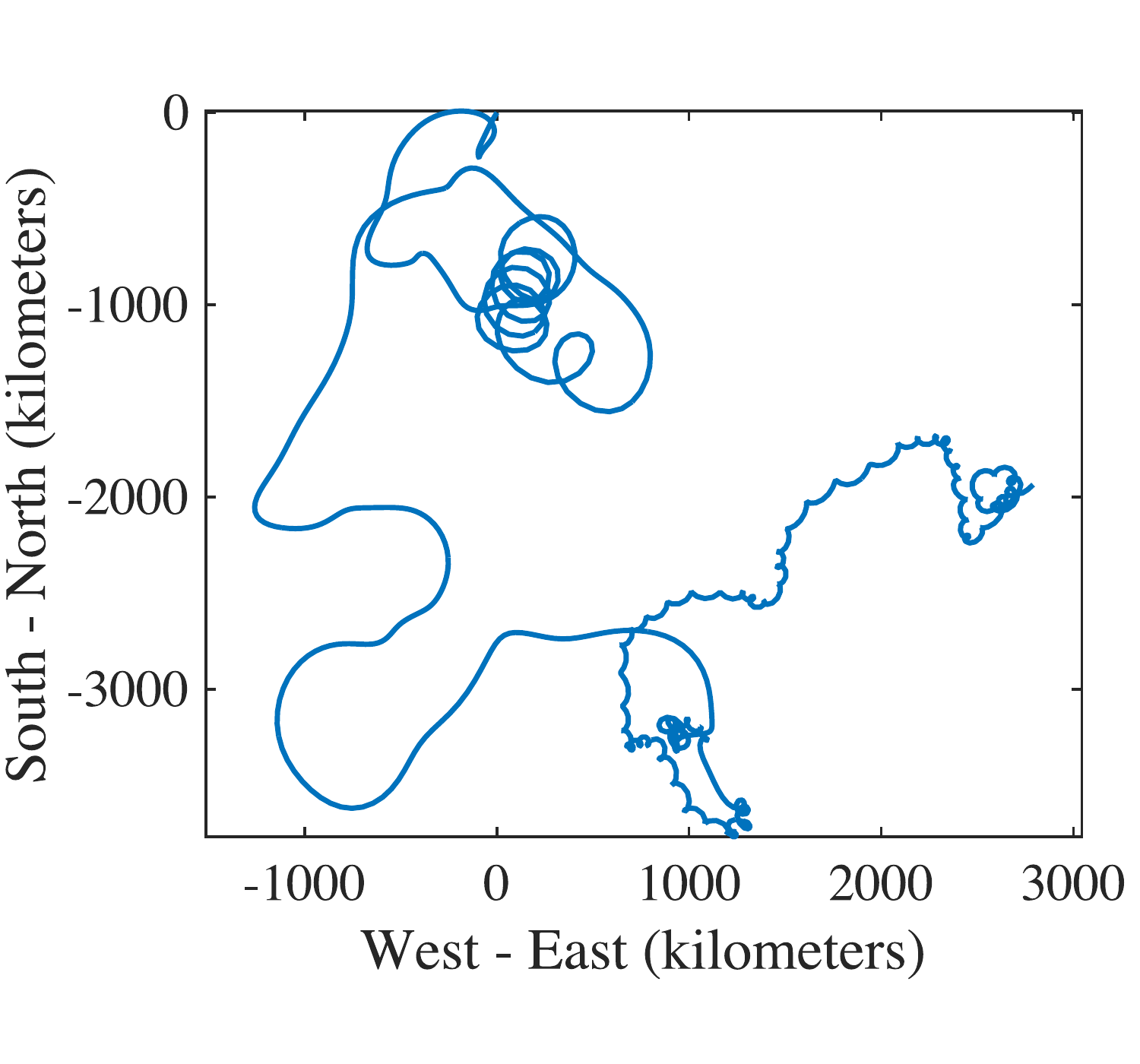}
\vspace{-4mm}
\caption{} \label{fig:1b}
\end{subfigure}
\hspace*{-2mm} 
\vspace{-4mm}
\begin{subfigure}{0.32\textwidth}
\includegraphics[width=0.98\textwidth]{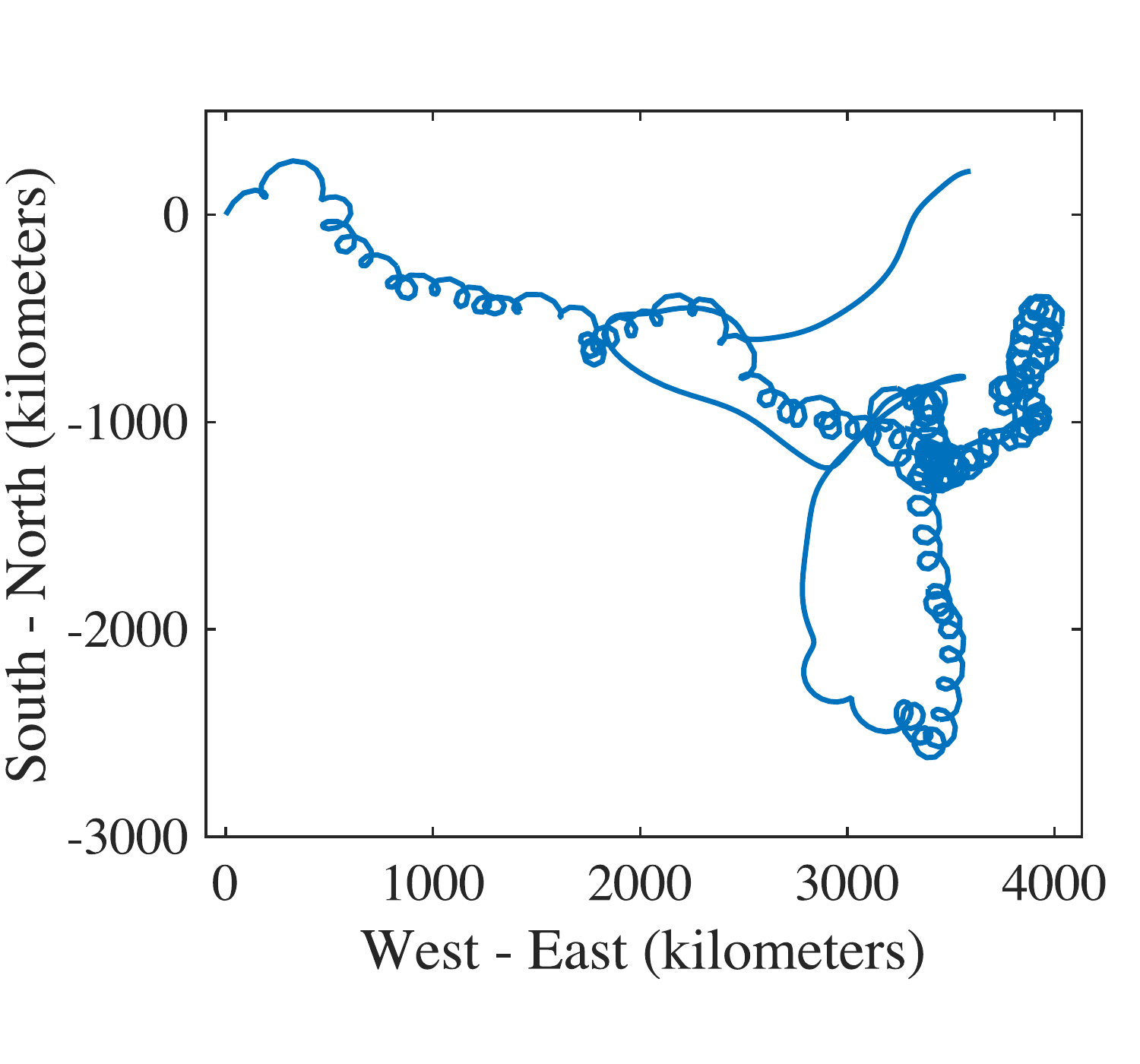}
\vspace{-4mm}
\caption{} \label{fig:1c}
\end{subfigure}
\vspace{4mm}
\caption{\label{fig:1} The trajectories of three different particles propagated in a large two-dimensional fluid turbulence simulation.} \label{drifterLoc}
\vspace{-7mm}
\end{figure*}

To develop our test procedure we draw inspiration from existing methods to detect nonstationarity in time signals. Here tests have been developed for real-valued data, focussing on different structural behaviour of the observed signal. Standard procedures are formulated in the frequency domain~\cite{priestley1969test,rao1973some,sakiyama2003testing,tsao1984tests} or the wavelet domain~\cite{nason2013test,neumann2000wavelet}. We propose to develop theory from the frequency domain, however, to capture nonlinear characteristics of the observed time signal, we develop a test in the time domain. Specifically our test procedure examines the power variance of the observed signal, i.e. the variability of the instantaneous variance over time.

This is, to our knowledge, the first test for nonstationarity designed specifically for complex-valued signals. We will demonstrate that testing the power variance can both identify {\em heteroscedastic} signals, where the variance of the signal changes over time, such as will seen shortly in Fig.~1(b), and signals with strong sinusoidal components that appear `phase-locked' as in Fig.~1(c). The latter is a strongly nonlinear feature and is associated with relatively low power variance, as we shall demonstrate. We will also derive the expectation of the test statistic under the null hypothesis, and show how the test can be efficiently performed in ${\cal O}\left(N\log N\right)$ time, where $N$ is the length of the observed signal, using the bootstrap and Fast Fourier Transforms.

\vspace{-1mm}
\section{Second-Order Properties and the Power Variance}\label{SecSec}
We first formulate theory for continuous signals, before detailing how approximations are made with discrete observations. Consider a continuous bivariate signal $\{x_t,y_t\}$ represented as a complex-valued quantity $z_t=x_t+iy_t$ at each time point $t$, where $i\equiv\sqrt{-1}$. If the complex-valued signal $\{z_t\}$ is second-order stationary then $\{z_t\}$ has a finite constant mean, and an autocovariance sequence $\E(z_t z_{t-\tau}^\ast)$ that is finite and does not depend on $t$, where $z_t^\ast$ denotes the complex conjugate of $z_t$. To fully describe the properties of the complex-valued signal, we also need to describe its relation sequence \cite{schreier2010statistical}, given by $\E(z_t z_{t-\tau})$, which is also independent of time for a second-order stationary signal.

A second-order nonstationary signal does not satisfy one or more of the constraints of $\E(z_t)$, $\E(z_t z_{t-\tau}^\ast)$ and $\E(z_t z_{t-\tau})$ being independent of $t$. There are a multitude of ways that this can happen. We focus on the simple case where $\var(z_t)\equiv E(|z_t|^2)$ varies with time, such that the signal is heteroscedastic. One simple class of such models is that of {\em uniformly modulated processes}~\cite{priestley1988non}, where $z_t=\sigma_t u_t$ where $u_t$ is a stationary process, and $\sigma_t$ is a deterministic signal. Hence, a measure of nonstationarity during the time interval $(0,T)$ is the time mean of the squared deviation of the variance from its time-mean value, which we term the {\em power variance}, given by
\begin{equation}
\Omega(T)=\frac{1}{T}\int_0^T \left(\var \{ z_t\}- \bar \sigma^2\right)^2\,dt,
\label{eq1}
\end{equation}
where the {\em sample variance}, or {\em average power}, is given by
\begin{equation}
\bar \sigma^2=\frac{1}{T}\int_0^T \var \{ z_t\}\;dt.
\end{equation}
It is clear that for a stationary signal, the variance of $z_t$ is constant over time such that $\Omega(T)\equiv 0$. However for nonstationary signals it would commonly be the case that $\Omega(T)\neq 0$, depending on the form of nonstationarity.

If we now consider discrete observations of $\{z_t\}$, then we do not observe $ \var \{ z_t\}$ directly, so instead we calculate a test statistic. For a  regularly sampled complex-valued signal, $\mathbf{z}=\left(z_0,z_1,z_2,...,z_{N-1}\right)$, we define the {\em observed power variance} as
\begin{equation}
\widehat \Omega({\bf z})=\frac{1}{N}\sum_{n=0}^{N-1} \left( \left|z_n\right|^2-\widehat {\sigma}^2\right)^2,
\label{eq3}
\end{equation}
where
\begin{equation}
\widehat {\sigma}^2=\frac{1}{N}\sum_{n=0}^{N-1} \left|z_n\right|^2.
\label{eq4}
\end{equation}
For a discretely observed stationary signal, it is {\em not} typically the case that $\widehat\Omega({\bf z})=0$, in contrast to $\Omega(T)$ in (\ref{eq1}). This is because of the natural stochasticity in $z_t$, which yields $|z_n|^2\neq\widehat\sigma^2$ at each observed time point. It is therefore possible for a nonstationary signal to have a smaller observed power variance than a stationary signal---a feature which we account for in our algorithm by performing two-sided statistical tests.

\vspace{-1mm}
\section{Bootstrap Test Procedure}\label{bootSec}
To test whether an observed signal is stationary or not, we first generate a number of randomised signal replicates that are constructed to be stationary, where each replicate has the same overall variability as the observed signal. We then contrast the power variance of the observed signal and that of the replicates. If the power variance of the observed signal is significantly different from the typical power variance found in the replicates, then we can reject the null hypothesis of stationarity, and identify the signal as nonstationary.

We generate the signal replicates using the bootstrap method~\cite{efron1979bootstrap}. We do this in the frequency domain by following the exact procedure of~\cite{theiler1992testing}. This simple procedure takes any complex-valued signal and calculates its Discrete Fourier Transform (DFT) given by
\begin{equation}
\left|Z_k\right|e^{i\phi_k}=\sum_{n=0}^{N-1}z_n e^{-2\pi ikn/N}.\label{fourDecomAmps}
\end{equation}
The Discrete Fourier Transform is complex-valued and we have decomposed this in terms of an amplitude $|Z_k|$ and corresponding phase $\phi_k$ at each frequency $k$. The idea of \cite{theiler1992testing} is to fix the amplitudes $|Z_k|$ but uniformly generate randomised phases, $\widetilde{\phi}_k\sim\mathcal{U}(-\pi,\pi)$, and then perform the inverse DFT into the time domain to generate a randomised replicate of the signal ${\bf\widetilde{z}}=\left(\widetilde{z}_0,\widetilde{z}_1,\widetilde{z}_2,...,\widetilde{z}_{N-1}\right)$ given by
\begin{equation}
\widetilde{z}_n=\frac{1}{N}\sum_{k=0}^{N-1}\left|Z_k\right|e^{i\widetilde{\phi}_k}e^{2\pi ikn/N}.\label{permTimeSeries}
\end{equation}
By definition, the frequency components of a stationary signal have uncorrelated phases, and hence the replicated signal ${\bf \widetilde{z}}$ (which has completely random phases) is stationary, even if the observed signal ${\bf z}$ is not. See also \cite{theiler1992testing} in relation to this.

Our test procedure generates many such replicated signals and compares the power variance with that of the observed signal, to see if there is a significant difference. We detail the exact test procedure in Algorithm 1, where FFT denotes the Fast Fourier Transform and $\mathbb{I}(\cdot)$ denotes the indicator function. If the null is rejected then there is sufficient evidence to suggest that the observed signal is nonstationary, otherwise the null should not be rejected using this test procedure. Note that this is a two-sided test procedure---we have tested whether the observed power variance is significantly lower or higher than that found in the distribution of stationary replicates. One-sided tests can be performed by using the values of $q({\bf z})$ or $r({\bf z})$ in Algorithm 1, instead of $p({\bf z})$, to respectively test for significantly high or low power variance.

\begin{algorithm}
\caption{\label{mainAlgorithm}Bootstrap Power Variance Test}
\begin{algorithmic}
\STATE {\bf INPUTS:} ${\bf z}=(z_0,\ldots,z_{N-1})$, $B$
\item $\widehat\Omega({\bf z})\leftarrow\frac{1}{N}\sum_{n=0}^{N-1} \left( \left|z_n\right|^2-\frac{1}{N}\sum_{n=0}^{N-1} \left|z_n\right|^2\right)^2$
\item $\{Z_k\}\leftarrow {\rm FFT}({\bf z})$
\FOR{$b=1$ to $B$}
\STATE $\widetilde{\phi}_k\overset{iid}{\sim}\mathcal{U}(-\pi,\pi)$, for $k=0,...,N-1$
\item${\bf \widetilde{z}}^{(b)}\leftarrow\frac{1}{N}{\rm FFT}\left(\left\{\left|Z_k\right|e^{i\widetilde{\phi}_k}\right\}\right)$
\item$\widehat\Omega\left({\bf \widetilde{z}}^{(b)}\right)\leftarrow\frac{1}{N}\sum_{n=0}^{N-1} \left( \left|\widetilde{z}^{(b)}_n\right|^2-\frac{1}{N}\sum_{n=0}^{N-1} \left|\widetilde{z}^{(b)}_n\right|^2\right)^2$
\ENDFOR
\STATE $q({\bf z})\leftarrow\frac{1}{B}\sum_{b=1}^{B}\mathbb{I}[\widehat\Omega\left({\bf \widetilde{z}}^{(b)}\right)>\widehat\Omega({\bf z})]$
\STATE $r({\bf z})\leftarrow\frac{1}{B}\sum_{b=1}^{B}\mathbb{I}[\widehat\Omega\left({\bf \widetilde{z}}^{(b)}\right)<\widehat\Omega({\bf z})]$
\STATE $p({\bf z}) \leftarrow 2\min\left\{q({\bf z}),r({\bf z})\right\}$
\IF {$p({\bf z})<0.05$}
\STATE Reject null
\ENDIF 
\end{algorithmic}
\end{algorithm}

By using the FFT to carry out the Fourier transforms, we can achieve $\mathcal{O}\left(N\text{log}N\right)$ computational cost, for each bootstrap iteration. For example, performing a test on one complex-valued signal with $N=1000$ and $B=1000$ on a 2.6 GHz 2012 Macbook Pro takes 0.072s. As the bootstraps are generated independently, the method is straightforward to run in parallel with larger datasets. We note that bootstrapping, phase scrambling, and similar procedures for generating time series replicates are discussed further in~\cite{shinozuka1972digital,shinozuka1991simulation,percival1993simulating,kirch2011tft}.

\vspace{-1mm}
\section{Illustrative Examples}\label{illEgSect}
By way of illustrative examples, we present three canonical cases for investigating our test procedure. All code in this paper is generated in Matlab\textsuperscript{\textregistered} and is available online from \url{http://www.ucl.ac.uk/statistics/research/spg/software}, and all results are exactly reproducible. Also free to download is an algorithm for applying the test procedure to any observed complex-valued signal. 

The first example we present is an autoregressive AR(1) model, where the sequence $\{z_n\}$ is defined by
\begin{align*}
x_n&=0.9x_{n-1}+0.1\sigma^{(x)}_n,\enspace\text{where}\enspace\sigma^{(x)}_n\overset{iid}{\sim}\mathcal{N}(0,1),\\
y_n&=0.9y_{n-1}+0.1\sigma^{(y)}_n,\enspace\text{where}\enspace\sigma^{(y)}_n\overset{iid}{\sim}\mathcal{N}(0,1),\\
z_n&=\frac{x_n+iy_n}{\sqrt{2}}.
\end{align*}
\noindent
 \begin{wrapfigure}{r}{0.56\columnwidth}
\vspace{-4ex}
\hspace{-3.5ex}\includegraphics[width=0.56\columnwidth]{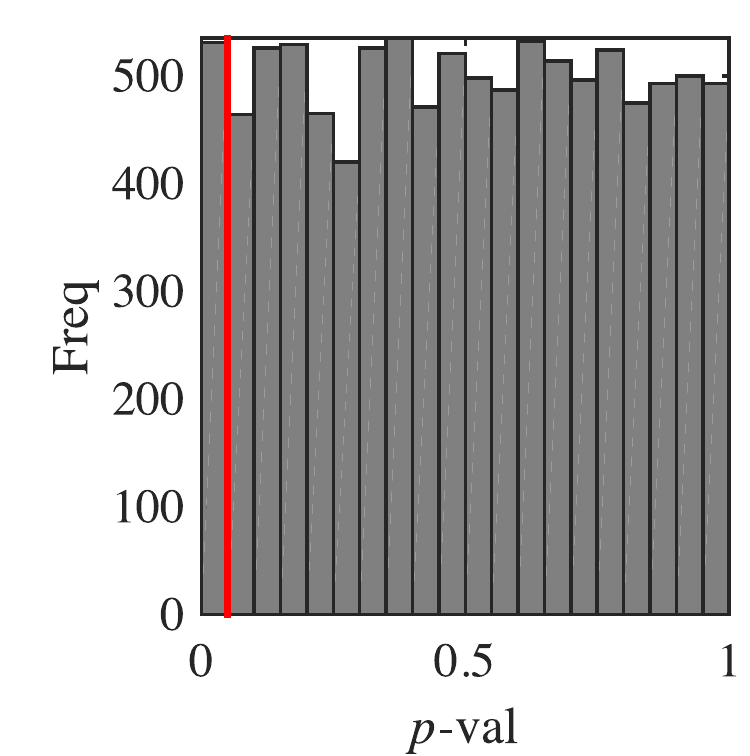}
\vspace{-1ex}
\caption{Distribution of {\it p}-values calculated from $p({\bf z})$ in Algorithm \ref{mainAlgorithm} for AR(1) process.}\label{ar1eg}
\vspace{-4ex}
\end{wrapfigure}
This model generates stationary signals. We now check if our test procedure correctly does not reject the null for realisations from this model. We set $N=1000$ and $B=1000$ in Algorithm~\ref{mainAlgorithm} and compute $p({\bf z})$ for a signal generated from this model. We perform a Monte Carlo simulation and repeat this procedure 10,000  times with independently generated signals from the AR(1) model. We then get a distribution of \textit{p}-values which are shown in Fig.~\ref{ar1eg}, with the 5\% significance level indicated with a vertical red line. It is clear that there is no tendency to reject the null for this process, with 5.21\% of null hypotheses rejected, which is consistent with a type I error level of 5\%.

\begin{wrapfigure}{r}{0.52\columnwidth}
\vspace{-3ex}\hspace{-3.5ex}\includegraphics[width=0.56\columnwidth]{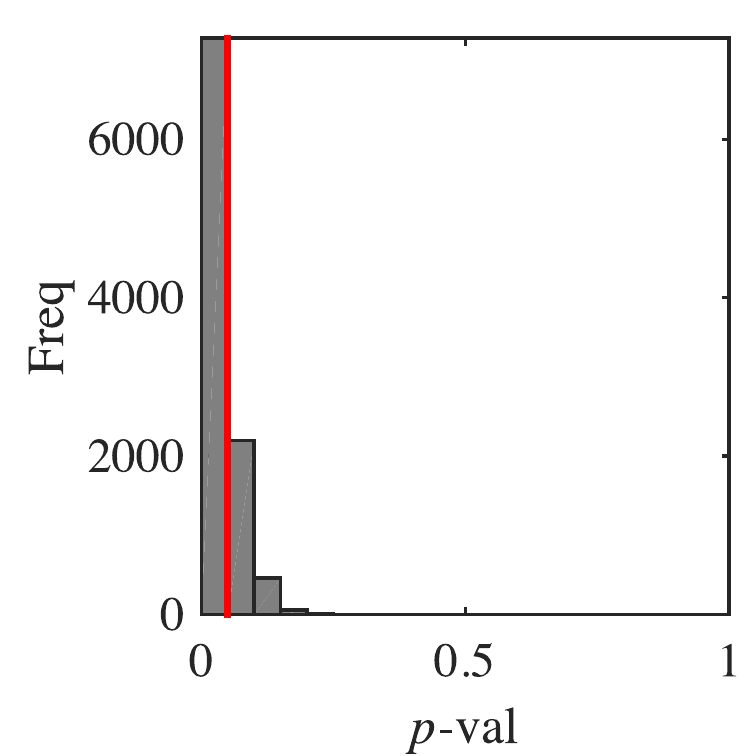}
\vspace{-1ex}
\caption{Jump process with {\it p}-values calculated from $q({\bf z})$ in Algorithm \ref{mainAlgorithm}.}\label{jpEg}
\vspace{-4ex}
\end{wrapfigure}

The second canonical example is a basic jump process, with $\{z_n\}$ defined by
\begin{align*}
x_n&=\sigma^{(x)}_n,\enspace\text{where}\enspace\sigma^{(x)}_n\overset{iid}{\sim}\mathcal{N}(0,1),\\
y_n&=\sigma^{(y)}_n,\enspace\text{where}\enspace\sigma^{(y)}_n\overset{iid}{\sim}\mathcal{N}(0,1),\\
z_n&=\begin{cases}
    1+(x_n+iy_n)/\sqrt{2},\text{ if }n\leq N/2,\\
    3+(x_n+iy_n)/\sqrt{2},\text{ if }n>N/2.
\end{cases}
\end{align*}
\noindent
This model generates nonstationary signals with a significant change-point midway through the series. We check if our test procedure correctly rejects the null for realisations from this model. Again setting $N=1000$ and $B=1000$, we generate 10,000 independent instantiations of such a jump process, and compute \textit{p}-values using Algorithm~\ref{mainAlgorithm}.  This time we perform a one-sided test and report values for $q({\bf z})$ in Algorithm~\ref{mainAlgorithm} to check for high power variance. The one-tailed test \textit{p}-values are shown in Figure \ref{jpEg}, with the 5\% significance level similarly indicated with a vertical red line. In this Monte Carlo study, there is a tendency to reject the null for this process, with an alternative of increased power variance: 71.8\% of null hypotheses are rejected at the 5\% significance level. This occurs because the process is nonstationary and has a significant change-point, which stationary replicates will not have. This yields a significantly higher observed power variance in most cases. The null is not always rejected however as the signal-to-noise ratio is set reasonably low.

\begin{wrapfigure}{r}{0.52\columnwidth}
\vspace{-4ex}
\hspace{-3.5ex}\includegraphics[width=0.56\columnwidth]{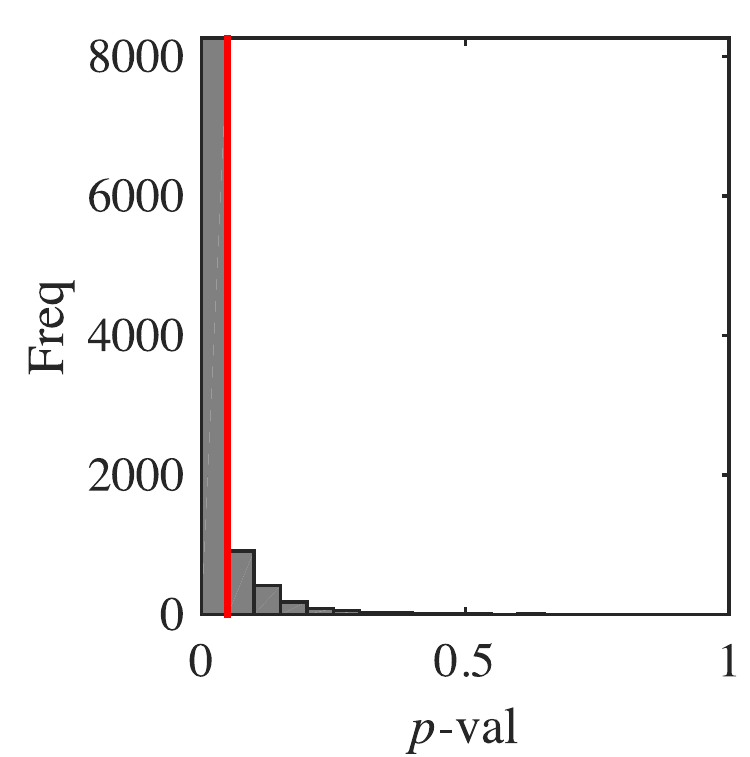}
\vspace{-1ex}
\caption{Cyclo-stationary process with {\it p}-values calculated from $r({\bf z})$ in Algorithm \ref{mainAlgorithm}.}\label{swEg}
\vspace{-7ex}
\end{wrapfigure}

The final illustrative example is a cyclo-stationary process, with $\{z_n\}$ defined by
\begin{equation*}
z_n=ae^{i\omega n/N}+\frac{\sigma^{(x)}_n+i\sigma^{(y)}_n}{\sqrt{2}},\enspace\text{where}\enspace\sigma^{(x)}_n,\sigma^{(y)}_n\overset{iid}{\sim}\mathcal{N}(0,1).
\end{equation*}
We set $\omega=10$ and $a=1$ such that the signal-to-noise ratio is relatively low. This model generates nonstationary signals because the sinusoid is `phase-locked' over time. For our simulations we again set
$N=1000$ and $B=1000$, and perform 10,000 Monte Carlo replicates. We perform a one-sided test and report values for $r({\bf z})$ in Algorithm~\ref{mainAlgorithm} to check for low power variance. The one-tailed test {\it p}-values are shown in Figure \ref{swEg}, with the 5\% significance level again indicated with a vertical red line. It is clear that there is a tendency to reject the null for this process, with an alternative of decreased power-variance: 82.3\% of null hypotheses are rejected at the 5\% significance level. We observe a decreased power-variance here because the fixed phase of the sinuoid implies that the amplitude of the signal will be less variable than the stationary replicates---which will not be `phase-locked' and exhibit more variability in terms of their instantaneous variance.

The illustrative examples given above (Figures \ref{ar1eg}-\ref{swEg}) are all based on signal length $N=1000$. Performance decreases as signal length decreases, however the null is still rejected in many more cases in the nonstationary examples than would be expected for a stationary signal, for even the shortest signals. Table \ref{sigSummary} summarises these findings.

\begin{table}[H]
\vspace{7ex}
\begin{center}
{\small
\begin{tabular}{|c|c|c|c|}
  \hline
$N$ & AR & Jump & CS \\
  \hline
1000 & 5.21\% & 71.8\% & 82.3\% \\
500 & 4.81\% & 57.2\% & 60.6\% \\
200 & 5.11\% & 39.2\% & 36.1\% \\
100 & 5.16\% & 29.0\% & 24.5\% \\
50 & 5.51\% & 21.5\% & 17.1\% \\
20 & 5.94\% & 14.8\% & 11.7\% \\
10 & 5.53\% & 11.5\% & 9.50\% \\
  \hline
\end{tabular}
}
\vspace{1mm}
\caption{Percentages of null hypotheses rejected at the 5\% level for various signal lengths $N$, for auto-regressive (AR), jump and cyclo-stationary (CS) processes. All settings except $N$ are identical to those used to generate Figures \ref{ar1eg}-\ref{swEg}.}\label{sigSummary}
\vspace{-7mm}
\end{center}
\end{table}

\section{Theory}\label{theorySection}
The bootstrap signal replicates, ${\bf \widetilde{z}}$, generated in Algorithm~\ref{mainAlgorithm} are used to approximate a distribution of power variances, under the null hypothesis that the observed signal is stationary. In this section we derive the expectation of this distribution theoretically. We denote $\E(\widehat\Omega({\bf \widetilde{z}}))$ as this expectation and analytically derive its form in terms of the Fourier amplitudes $|Z_k|$ found in (\ref{fourDecomAmps}).
\begin{Proposition}
For a given complex-valued signal ${\bf z}$ then the power variance of bootstrap replicates $\bf{\widetilde{z}}$ generated using (\ref{permTimeSeries}) has expectation given by
\begin{equation}
\E(\widehat\Omega({\bf \widetilde{z}}))=\frac{1}{N^4}\left\{\left(\sum_{k=0}^{N-1}\left|Z_k\right|^2\right)^2-\sum_{k=0}^{N-1}\left|Z_k\right|^4\right\},\label{powerVarDef}
\end{equation}
conditional on the observed signal $\mathbf{z}=\left(z_0,z_1,z_2,...,z_{N-1}\right)$, where $|Z_k|$ is found using (\ref{fourDecomAmps}).
\end{Proposition}

This proposition is particularly useful when performing one-sided tests, in which one is looking for evidence that $\widehat\Omega({\bf z})$ is either unreasonably high or low. This is because the expectation can be pre-computed before implementing the bootstrap. Then if the expectation is greater than the observed power variance when testing for a high power variance---or vice-versa for a low power variance---there is no need to perform the bootstrap as the null can already conclusively not be rejected at this stage. The proof of the proposition is as follows.

\begin{proof}
First we directly compute from (\ref{eq3}) and (\ref{eq4}) that the expectation for the replicates takes the form
\begin{align}
\E(\widehat\Omega({\bf \widetilde{z}}))&=\E\left[\frac{1}{N}\sum_{n=0}^{N-1}\left|\widetilde{z}_n\right|^4-\left(\frac{1}{N}\sum_{n=0}^{N-1}\left|\widetilde{z}_n\right|^2\right)^2\right]\nonumber\\
&=\E\left(\left|\widetilde{z}_n\right|^4\right)-\E\left[\left(\frac{1}{N}\sum_{n=0}^{N-1}\left|\widetilde{z}_n\right|^2\right)^2\right]\label{proof1}.
\end{align}
Examining the first term of (\ref{proof1}) and using (\ref{permTimeSeries}) we see that
\begin{align}
\E\left(\left|\widetilde{z}_n\right|^4\right)=&\frac{1}{N^4}\E\left(\sum_{k=0}^{N-1}\sum_{l=0}^{N-1}\sum_{p=0}^{N-1}\sum_{q=0}^{N-1}\left|Z_k\right|\left|Z_l\right|\left|Z_p\right|\left|Z_q\right|\right.\times\nonumber\\
&\quad\quad\quad\quad\left.e^{2\pi i(k+l-p-q)n/N}e^{i\left(\widetilde{\phi}_k+\widetilde{\phi}_l-\widetilde{\phi}_p-\widetilde{\phi}_q\right)}\right)\nonumber\\
=&\frac{1}{N^4}\sum_{k=0}^{N-1}\left|Z_k\right|^4+\frac{2}{N^4}\sum_{k=0}^{N-1}\sum_{l\neq k}\left|Z_k\right|^2\left|Z_l\right|^2,\label{proof2}
\end{align}
in which the Fourier amplitude coefficients $|Z_k|$ are regarded as deterministic. This last expression follows as the phases are generated randomly meaning
\begin{align*}
\E\left[e^{i\left(\widetilde{\phi}_k+\widetilde{\phi}_l-\widetilde{\phi}_p-\widetilde{\phi}_q\right)}\right]=&\begin{cases}
    1, & \text{if}\enskip p=k\text{ and }q=l,\\
    & \enskip\text{or if }q=k\text{ and }p=l,\\
    0, & \text{otherwise},
\end{cases}
\end{align*}
and thus the factor of two in the second term of (\ref{proof2}) follows from the two possibilities by which the indices in the summation can pair together ($p=k$ and $q=l$, or $q=k$ and $p=l$). The first term at the right-hand side of (\ref{proof2}) arises from the special case $q=p=l=k$. Equation (\ref{proof2}) can then be rewritten as
\begin{equation}
\E\left(\left|\widetilde{z}_n\right|^4\right)=\frac{2}{N^4}\left(\sum_{k=0}^{N-1}\left|Z_k\right|^2\right)^2-\frac{1}{N^4}\sum_{k=0}^{N-1}\left|Z_k\right|^4.\label{Ez4final}
\end{equation}
The second term of (\ref{proof1}) can be immediately found in terms of $|Z_k|$ using the Parseval-Rayleigh relationship
\begin{equation}
\E\left[\left(\frac{1}{N}\sum_{n=0}^{N-1}\left|\widetilde{z}_n\right|^2\right)^2\right]=\left(\frac{1}{N^2}\sum_{k=0}^{N-1}\left|Z_k\right|^2\right)^2.\label{proof3}
\end{equation}
The proposition then follows by substituting in (\ref{Ez4final}) and (\ref{proof3}) into (\ref{proof1}).
\end{proof}
The full analytic distribution of $\widehat\Omega({\bf \widetilde{z}})$ cannot be easily found for a given signal ${\bf z}$. Even if the observed signal is Gaussian, the distribution of power variances will in general not be Gaussian for finite $N$. For this reason we do not proceed with an analytic approach of performing the power variance test, and instead the bootstrap method of Section~\ref{bootSec} is preferred.

\section{Particle Trajectories in Two-Dimensional Turbulence}\label{turbTSsect}
\begin{figure*}[ht!]
\centering
\begin{subfigure}{0.32\textwidth}
\includegraphics[width=0.98\textwidth]{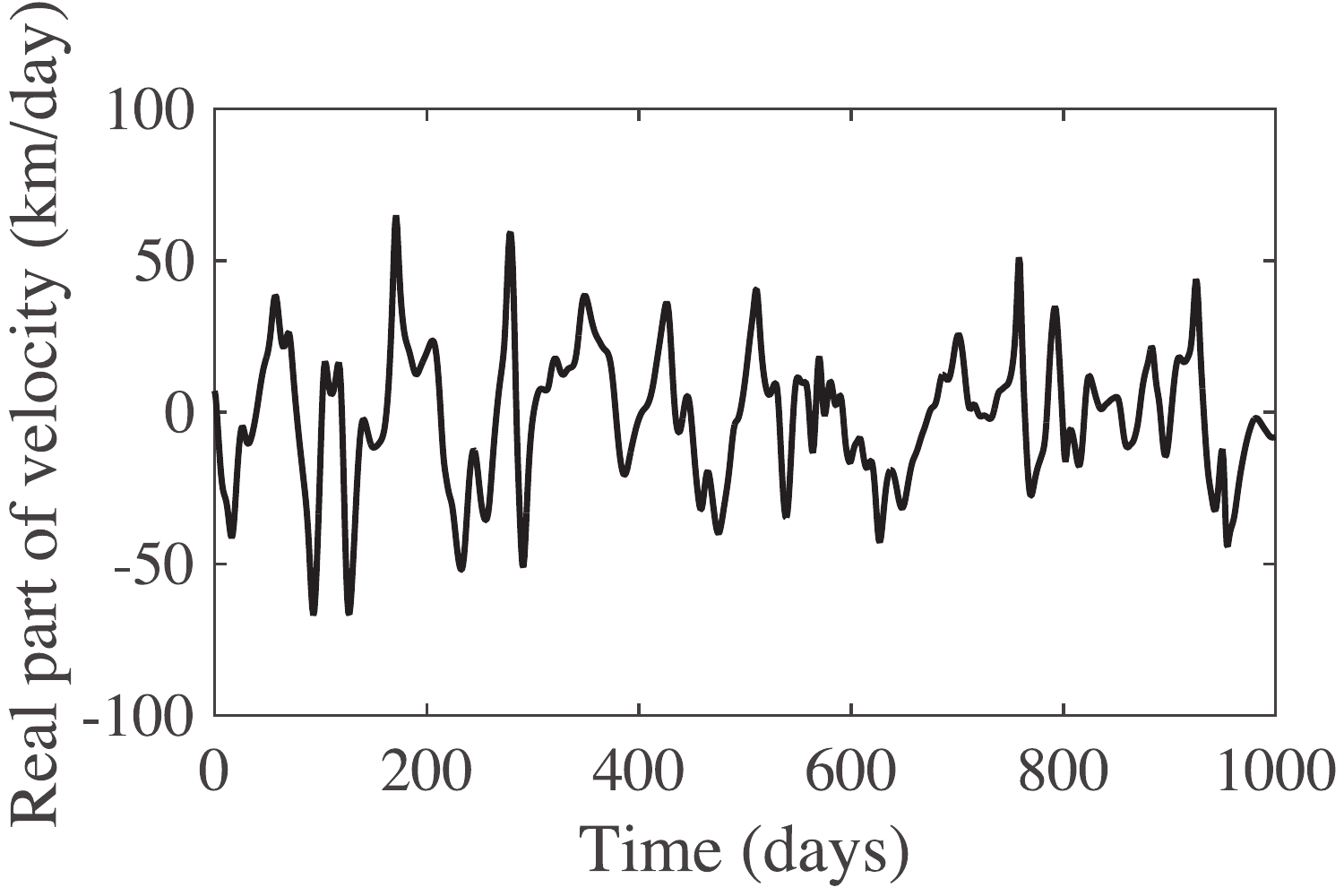}
\caption{}
\end{subfigure}
\hspace{1mm}
\begin{subfigure}{0.32\textwidth}
\includegraphics[width=0.98\textwidth]{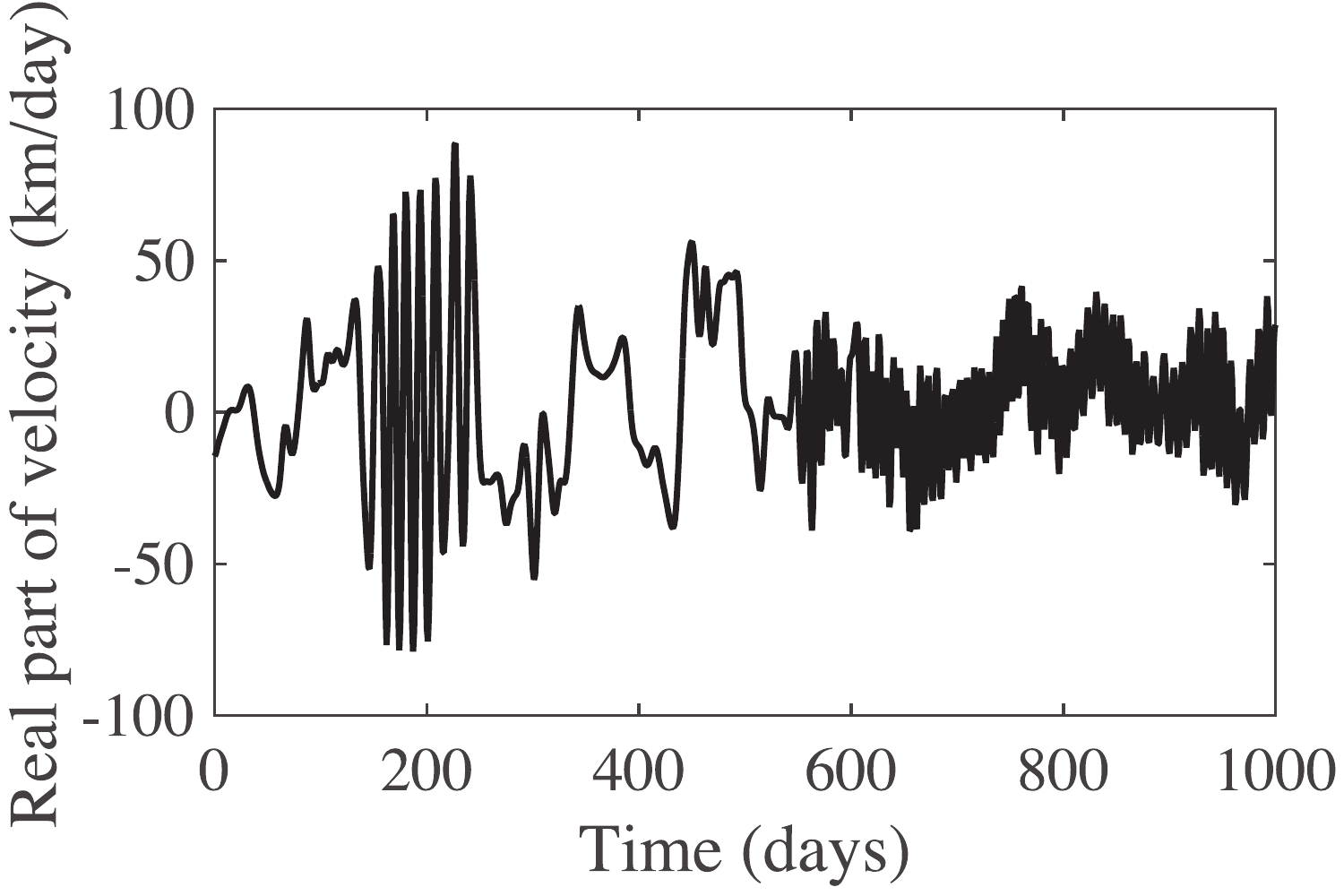}
\caption{}
\end{subfigure}
\hspace{1mm}
\begin{subfigure}{0.32\textwidth}
\includegraphics[width=0.98\textwidth]{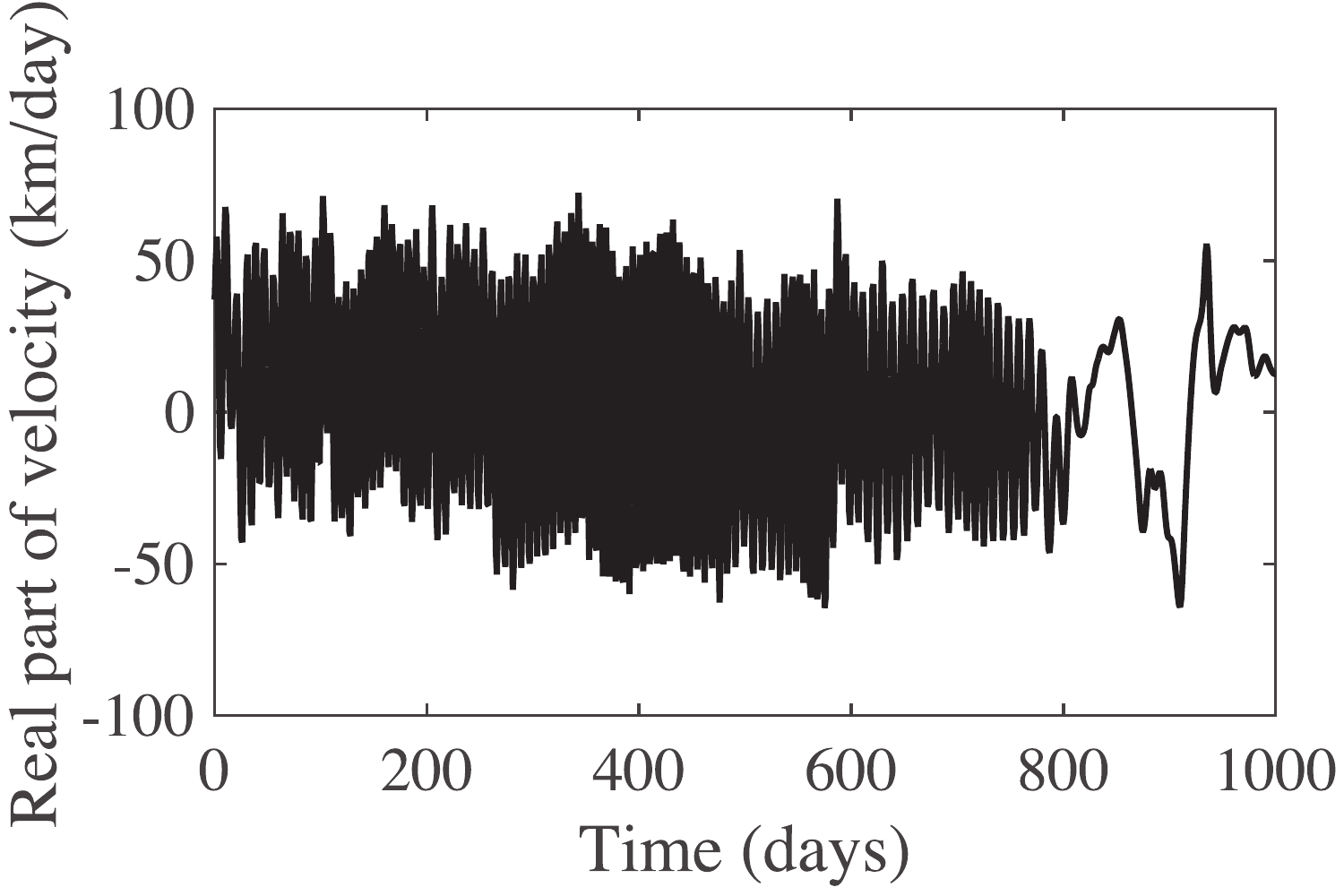}
\caption{}
\end{subfigure}
\vspace{-2mm}
\caption{The real part of the velocity signal from the trajectories plotted in Fig. 1. The time length of the signal is 1000 days. The imaginary part of the velocity signal is similar in shape and magnitude, but the rapid oscillations seen in the real part typically coincide with oscillations in the imaginary part that are ninety degrees out of phase, leading to looping motions on the complex plane, which can be seen in Fig. 1.} \label{realTS}
\end{figure*}
\begin{figure*}[ht!]
\centering
\begin{subfigure}{0.32\textwidth}
\includegraphics[width=0.99\textwidth]{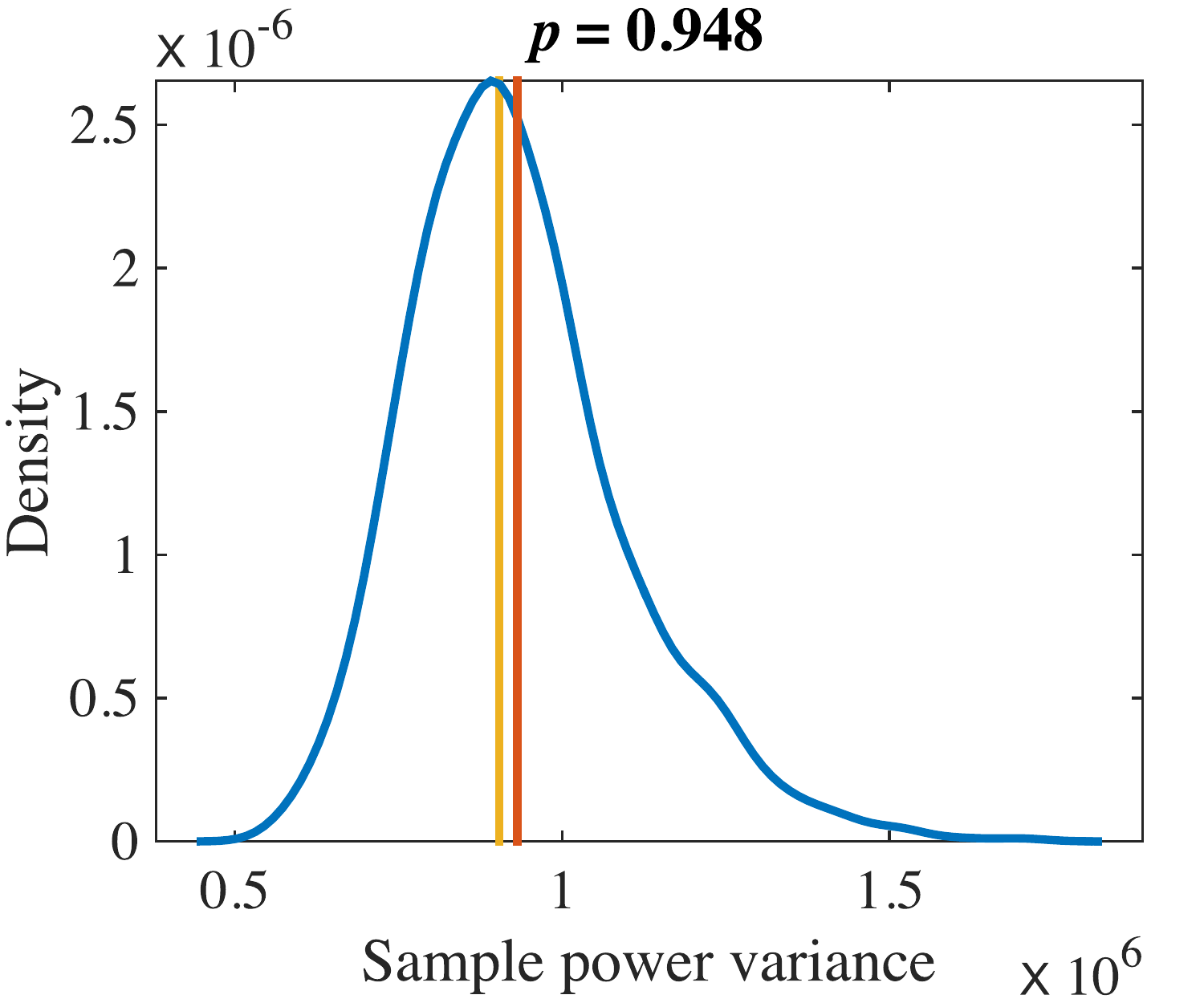}
\caption{}
\end{subfigure}
\begin{subfigure}{0.32\textwidth}
\includegraphics[width=0.99\textwidth]{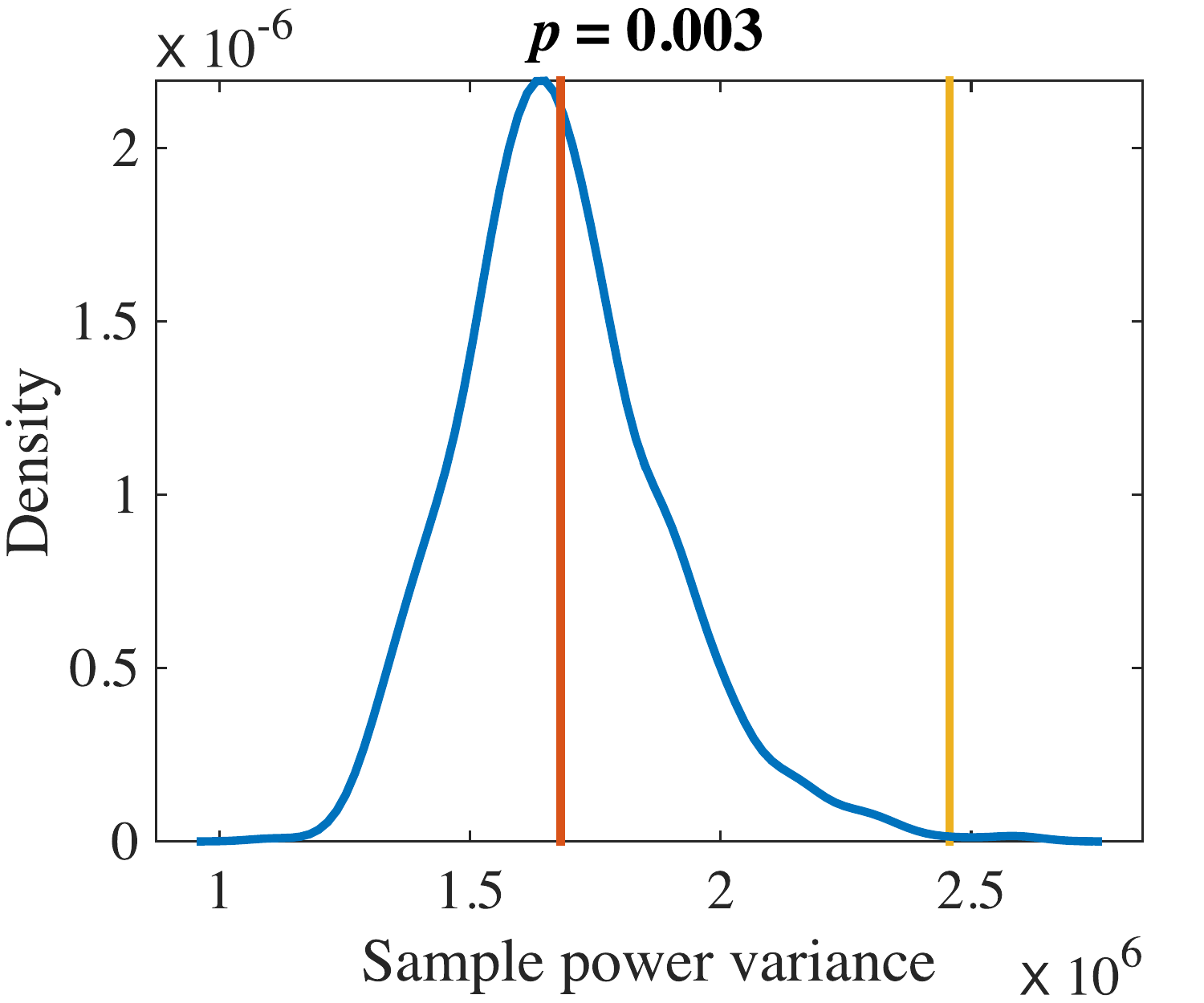}
\caption{}
\end{subfigure}
\begin{subfigure}{0.32\textwidth}
\includegraphics[width=0.99\textwidth]{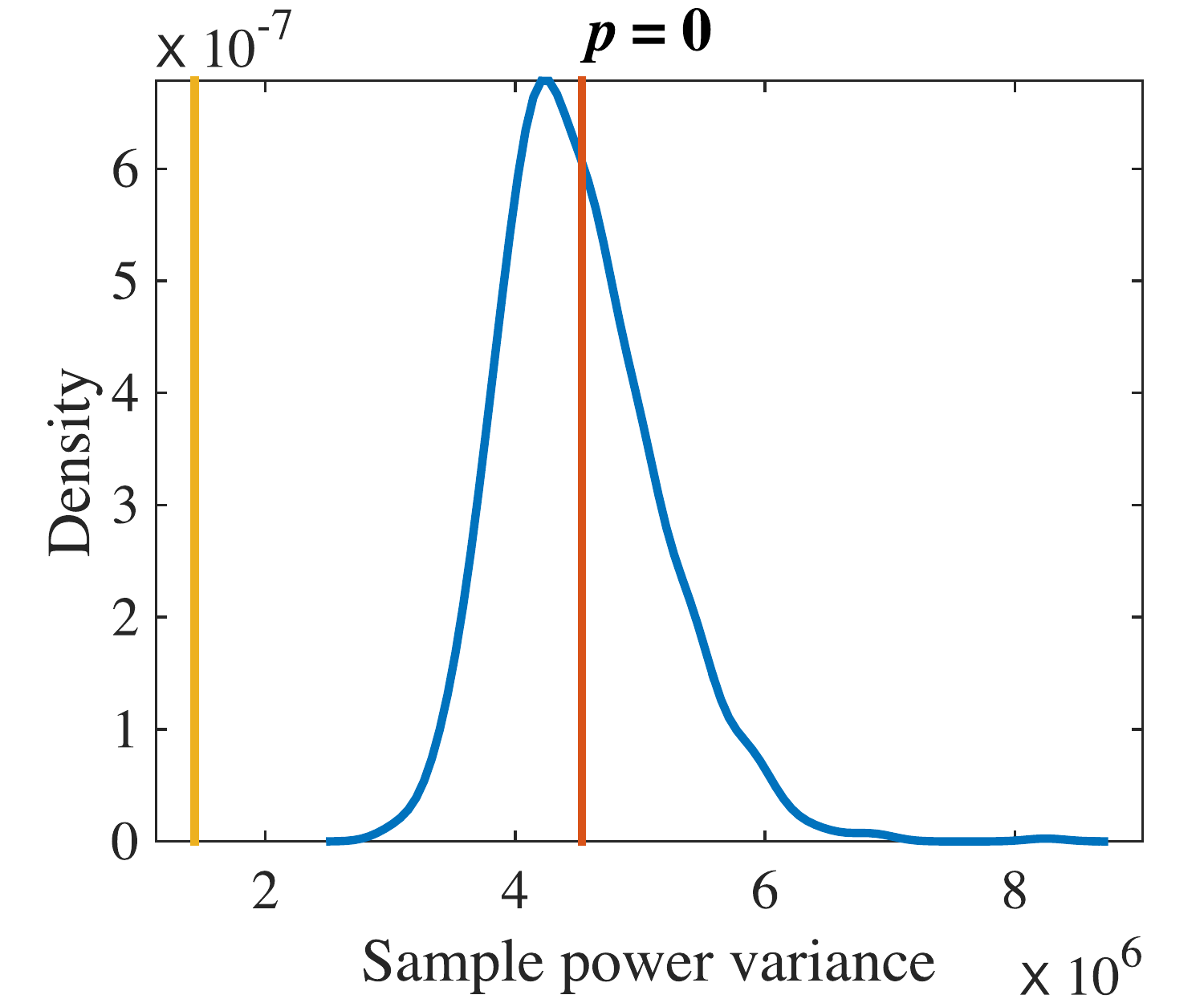}
\caption{}
\end{subfigure}
\vspace{-2mm}
\caption{Results from bootstrap power variance tests for the trajectories plotted in Fig. 1. Each plot shows the observed power variance (\textcolor{orange}{$\boldsymbol{|}$}), as well as the distribution of power variances from the bootstrap replicates (\textcolor{blue}{|}). The theoretical expectation of this distribution is also given (\textcolor{red}{$\boldsymbol{|}$}). The \textit{p}-value of the observed power variance is reported above the figure, and values less than 0.05 lead to rejection of the null hypothesis that the signal is stationary.} \label{hypothesisTests}
\vspace{-1mm}
\end{figure*}

Quasigeostrophic (QG) turbulence \cite{charney1971geostrophic} is a model for large-scale fluid motions in the ocean and atmosphere that is predominantly two-dimensional and is dominated by a balance between the pressure force and the apparent Coriolis force. Qualitatively, the fluid can be divided into two regimes---regions of swirling masses of fluid known as eddies, and regions in between the eddies. The positions of particles transported by the flow, referred to as drifters, are tracked over time, leading to trajectory curves. In this simulation there are some drifter trajectories which experience nonstationarity due to crossing eddy boundaries, thereby transitioning between eddy interiors and the ambient fluid, and others which do not. This section uses our test procedure to classify the trajectories into different types. 

As in Section~\ref{illEgSect}, all code and data is available online at {\em http://www.ucl.ac.uk/statistics/research/spg/software} and all results are exactly reproducible. Furthermore, the QG turbulence simulation can also be reproduced using the software available at {\em jeffreyearly.com/numerical-models/.}

Fig. \ref{drifterLoc} shows three trajectories from the simulation. In Fig \ref{drifterLoc}(a), we see an example of a stationary trajectory, while Fig. \ref{drifterLoc}(b) shows an example of a nonstationary trajectory comprised of portions with multiple distinct flow characteristics, and Fig. \ref{drifterLoc}(c) is another nonstationary trajectory dominated by a single eddy. Fig. \ref{realTS} shows the same examples, with the real part of the drifter {\em velocities} plotted, and we can again observe the same characteristics of stationary and nonstationary behaviour. 

Fig. \ref{hypothesisTests} shows the results of applying the bootstrap power variance test of Algorithm \ref{mainAlgorithm} (with $B=1000$) to the complex-valued velocity signals. Included in the figure are the corresponding variances of the null distributions, along with the observed power variances calculated according to (\ref{eq3}), and the means of the null power-variance distributions calculated from (\ref{powerVarDef}). In Fig. \ref{hypothesisTests}(a), the observed power variance is in the middle of the null distribution (and hence the null is not rejected), because this signal appears reasonably stationary: the {\it p}-value here is calculated as $p({\bf z})$ in Algorithm~\ref{mainAlgorithm}. In Fig. \ref{hypothesisTests}(b), the observed power variance is in the extreme upper tail of the null distribution (leading to rejection of the null), because this signal is clearly nonstationary: the {\it p}-value here is calculated as $q({\bf z})$ in Algorithm~\ref{mainAlgorithm}. Finally in Fig. \ref{hypothesisTests}(c), the observed power variance is in the extreme lower tail of the null distribution (and again the null is rejected). This is because this signal is nonstationary in a different way: the sinusoidal oscillations are `phase-locked' in such a way that the instantaneous variance of the signal has {\em less} variability than stochastic stationary replicates. The {\it p}-value here is calculated as $r({\bf z})$ in Algorithm~\ref{mainAlgorithm}.
 
 \section{Conclusions}
In this paper, we have proposed an algorithm for testing for nonstationarity in complex-valued signals. The algorithm is based on calculating the power variance, a measure of the instantaneous variance of power of the observed signal. The algorithm can learn two different types of nonstationarity, this leading to a two-sided test. Both tails are indeed informative: if the power variance is too large, then this indicates that the variance of the signal is too variable to be classified as stationary. On the other hand, if the power variance is too small, then this indicates that the variance of the signal is {\em less} than that expected under the hypothesis of stationarity.

To perform the test, the algorithm employs the bootstrap method. The bootstrapped signals have the same overall variance, but are stationary by construction and therefore have constant variability. By comparing the original signal with these replicates we can determine if the original signal has the characteristics of a stationary signal. We do this by comparing the power variance. Significantly high power variance indicates a heteroscedastic process which is nonstationary. Significantly low power variance indicates a very small degree of variability; this is indicative of a strong `phase-locking' whereby various Fourier components interact to generate a relatively uniform level of signal power.

Developing fully assumption-free methods of detecting signal nonstationarity is of great current interest. Our focus in this paper has been on coupled pairs of signals; in general we would study signals of arbitrary dimensionality. In such settings it is significantly harder to determine natural test statistics, and extending our methods to such settings remains an outstanding challenge. Finally, while the focus of this paper has been on naturally complex-valued signals, the same basic approach should apply to real-valued signals that have been made complex-valued though taking the analytic part \cite{marple1999computing}, which is a promising extension for future work.

\section*{Acknowledgments}
TEB, AMS and SCO acknowledge funding from EPSRC grant no. EP/L025744/1. SCO also acknowledges funding from EPSRC grant no. EP/I005250/1. The research of AMS was supported by a Marie Curie International Outgoing Fellowship within the seventh European Community framework programme. The work of JML and JJE was supported by award \#1235310 from the Physical Oceanography Program of the United States National Science Foundation.



\bibliographystyle{IEEEtran}
\bibliography{references}
%
%
%
%

\end{document}